%% file: 99-main.tex
\documentclass[sigconf,screen]{acmart}
\AtBeginDocument{%
  \providecommand\BibTeX{{%
    \normalfont B\kern-0.5em{\scshape i\kern-0.25em b}\kern-0.8em\TeX}}}

\copyrightyear{2021} 
\acmYear{2021} 
\setcopyright{acmlicensed}\acmConference[CIKM '21]{Proceedings of the 30th ACM International Conference on Information and Knowledge Management}{November 1--5, 2021}{Virtual Event, QLD, Australia}
\acmBooktitle{Proceedings of the 30th ACM International Conference on Information and Knowledge Management (CIKM '21), November 1--5, 2021, Virtual Event, QLD, Australia}
\acmPrice{15.00}
\acmDOI{10.1145/3459637.3482099}
\acmISBN{978-1-4503-8446-9/21/11}

\include{99-macros}

\begin{document}
\fancyhead{}

\title{Evaluating Fairness in Argument Retrieval}


\author{Sachin Pathiyan Cherumanal}
\orcid{https://orcid.org/0000-0001-9982-3944} 
\affiliation{
\institution{RMIT University}
\city{Melbourne}
\country{Australia}
}
\email{sachin.pathiyan.cherumanal@student.rmit.edu.au}

\author{Damiano Spina}
\orcid{https://orcid.org/0000-0001-9913-433X}
\affiliation{
\institution{RMIT University}
\city{Melbourne}
\country{Australia}
}
\email{damiano.spina@rmit.edu.au}

\author{Falk Scholer}
\orcid{https://orcid.org/0000-0001-9094-0810}
\affiliation{
\institution{RMIT University}
\city{Melbourne}
\country{Australia}
}
\email{falk.scholer@rmit.edu.au}

\author{W.~Bruce Croft}
\orcid{}
\affiliation{
\institution{University of Massachusetts}
\city{Amherst}
\state{MA}
\country{USA}
}
\email{croft@cs.umass.edu}


\include{00-abstract}




\maketitle

\input{01-introduction}

\input{02-relatedwork}

\input{03-fairness}

\input{04-setup}
\input{05-results}
\input{06-conclusions}

\parindent0pt{
\paragraph*{\bf Acknowledgments}
{
\small
The authors thank the organizers of Touch{\'e}@CLEF 2020 for making the runs available. This work is partially supported by the \grantsponsor{ARC}{Australian Research Council}{https://www.arc.gov.au/} (\grantnum{ARC}{DE200100064}, \grantnum{ARC}{DP190101113}).
}
}

\balance
\bibliographystyle{ACM-Reference-Format}
\bibliography{99-references}


\end{document}

%% file: 99-macros.tex
\usepackage{xspace}
\usepackage{soul}
\usepackage{upgreek}
\usepackage{euscript}
\usepackage{booktabs}
\usepackage{siunitx}

\usepackage{adjustbox}
\usepackage{flushend}
\usepackage{balance}

\usepackage[normalem]{ulem}

\newcommand{\ndcg}{{\small \sf nDCG}\xspace}
\newcommand{\ndcgTopk}{{\small \sf nDCG@5}\xspace}
\newcommand{\ndcgTopkCap}{{\small \sf {\bf nDCG@5}}\xspace}

\newcommand{\alphandcg}{{\small \sf {$\alpha$}-nDCG}\xspace}

\newcommand{\alphandcgTopk}{{\small \sf {$\alpha$}-nDCG@5}\xspace}

\newcommand{\alphandcgTopkCap}{{\small \bf {$\boldsymbol\alpha$}-nDCG@5}\xspace}

\newcommand{\rNDTopk}{{\small \sf rND@k}\xspace}
\newcommand{\rKLTopk}{{\small \sf rKL@k}\xspace}
\newcommand{\rRDTopk}{{\small \sf rRD@k}\xspace}

\newcommand{\rND}{{\small \sf rND@5}\xspace}
\newcommand{\rKL}{{\small \sf rKL@5}\xspace}

\newcommand{\rKLCap}{{\small \bf rKL@5}\xspace}

\newcommand{\rRD}{{\small \sf rRD@5}\xspace}

\newcommand{\rNDoriginal}{{\small \sf rND}\xspace}
\newcommand{\rKLoriginal}{{\small \sf rKL}\xspace}
\newcommand{\rRDoriginal}{{\small \sf rRD}\xspace}

\newcommand{\pro}{{\small \sf PRO}\xspace}
\newcommand{\con}{{\small \sf CON}\xspace}
\newcommand{\touche}{Touch{\'e}\xspace}

%% file: 00-abstract.tex
\begin{abstract}
Existing commercial search engines often struggle to represent different perspectives of a search query. Argument retrieval systems address this limitation of search engines and provide both positive (\pro) and negative (\con) perspectives about a user's information need on a controversial topic (e.g., climate change).
The effectiveness of such argument retrieval systems is typically evaluated based on topical relevance 
and argument quality, without taking into account the often differing number of documents shown for the argument stances (\pro or \con). Therefore, systems may retrieve relevant passages, but with a biased exposure of arguments. In this work, we analyze a range of non-stochastic fairness-aware ranking and diversity metrics to evaluate the extent to which argument stances are fairly exposed in argument retrieval systems.

Using the official runs of the argument retrieval task \touche at CLEF 2020, as well as synthetic data to control the amount and order of argument stances in the rankings, we show that systems with the best effectiveness in terms of topical relevance are not necessarily the most fair or the most diverse in terms of argument stance. 
The relationships we found between (un)fairness and diversity metrics shed light on how to evaluate group fairness -- in addition to topical relevance -- in argument retrieval settings.
\end{abstract}

%% file: 01-introduction.tex
\section{Introduction}
\label{sec:introduction}

The top search results in web search engines can be riddled with populism, conspiracy theories, and one-sidedness, not reflecting effectively the argumentative landscape for controversial topics (e.g., climate change)~\cite{potthast2019argument}.
Recently, 
argument retrieval systems such as \emph{args.me}~\cite{stein:2017r}
have been proposed. Argument retrieval systems aim to provide documents that cover both positive (\pro) and negative (\con) perspectives in search query results. 

While argument retrieval systems aim to tackle the problem of providing multiple perspectives on user queries, the arguments that are retrieved for a given topic may only cover one point of view. For example, a query such as \lq is universal basic income good\rq may return a list of relevant arguments supporting the statement but may expose fewer (or no) results with arguments against a universal basic income.
Even when all stakeholders are identified and their stances on a given argumentative topic are summarized~\cite{potthast2019argument}, the question arises whether rankings produced by such systems provide \emph{fair} exposure of the different relevant stances for a user's search on controversial topics \cite{kiesel2021meant}.
Moreover, there is no consensus on how \emph{fairness-aware argument retrieval} should be defined, and which metrics should be used to measure fairness in argument retrieval.

In this paper, we study the argument retrieval task -- where systems aim to provide assistance to users searching for relevant \pro and \con arguments on various societal topics -- proposed for the \touche Lab at CLEF 2020 \cite{bondarenko2020overview} as a fairness-aware ranking problem. We explore fairness metrics for evaluating argument retrieval systems by defining the protected groups for this particular problem.  Our results on the \touche submissions show that the retrieved results from an argument retrieval system are typically biased, and demonstrate that systems with the best effectiveness are not necessarily the most fair. Our analysis shows that fairness-aware rankings do not guarantee diverse results and vice-versa.

%% file: 02-relatedwork.tex
\section{Related Work}

\label{sec:Fairness_Intro}

\noindent \textbf{Fairness in Information Retrieval (IR).} Algorithm fairness has been studied across multiple disciplines and has attracted attention from researchers in IR \cite{allan2018swirl, castillo2018fairness, wilkie2014best}.  Most of these recent studies, regardless of the discipline \cite{dwork2012fairness,pedreshi2008discrimination,pedreschi2009measuring}, have considered \emph{individual fairness} and \emph{group fairness}. In this work we consider the group fairness scenario which requires that the demographics of those receiving a certain treatment is proportional to the demographics in the overall population -- also known as statistical parity \cite{dwork2011fairness}.
Most work on fairness in IR either measures the level of bias based on protected attributes or aims to mitigate such biases using fairness-aware algorithms or optimization techniques \cite{geyik2019fairness,singh2018fairness}. 
Our work is closely related to the former, as we measure group fairness in ranked outputs, but uses a slightly modified form of fairness measures that evaluate single-list rankings \cite{yang2017measuring}.
In a group fairness setting, a certain subset of the population is defined as a protected group. Most recent work uses characteristics such as race, gender, and disability status to define protected attributes, and therefore, protected groups \cite{zehlike2021fairness}.
For example, \citet{geyik2019fairness}~propose a fairness-aware ranking algorithm and fairness measures for talent search.
\citet{diaz2020evaluating} and \citet{asia2018equity} use protected attributes such as age or gender to analyze fairness across distributions of rankings, rather than in a single fixed ranking. 
Our focus is on evaluating group fairness for argument retrieval in a Cranfield paradigm offline evaluation setting, and we apply the notion of protected groups to analyze unbiased exposure of different argument stances.

\label{sec:Arg_Ret_Intro}
\noindent \textbf{Argument Retrieval.} 
An argument consists of a claim (e.g., ``We should abandon fossil fuels'') and a premise that justifies either attacking or supporting the claim \cite{dumani2020framework}. From an IR perspective, argument retrieval is the task of retrieving relevant supporting (\pro) and attacking (\con) justifications (premises) for a given query (claim). The motivation behind research in argument retrieval comes from the fact that search engines lack the capability to provide results that inform users about the premises and their stance (\pro or \con) towards the claim, as they currently ignore the argumentative nature of discussions in sources such as community question answering websites and debate portals \cite{bondarenko2020overview}. 
An example of an argument search engine is \textit{args.me} \cite{wachsmuth2017computational} which retrieves relevant arguments on a given (controversial) query from a focused collection of arguments crawled from a selection of debate portals.

%% file: 03-fairness.tex
\label{sec:methodology}
\section{Fairness and Diversity Metrics}

To study fairness in argument retrieval systems, we consider existing evaluation measures proposed to evaluate (un)fairness in rankings. As the task of balancing argument stances in a ranking can be seen as a particular case of search result diversification \cite{gao2020toward,draws2021assessing}, we also consider a commonly used diversity metric, \alphandcg~\citep{clarke2008novelty}.

\subsection{(Un)fairness Metrics}

We analyze three single-list metrics proposed by \citet{yang2017measuring}: normalized discounted difference (\rNDoriginal), normalized discounted KL-divergence (\rKLoriginal), and normalized discounted ratio (\rRDoriginal). We define these metrics using the notation described in Table~\ref{table:fairness_notations_table}, and explicitly indicate a cutoff $k$. Given that argument retrieval is a top-heavy ad-hoc retrieval task (e.g., the cutoff used at \touche challenge is $k=5$), we adapt set-based fairness at discrete points ranging from top-1 to top-5.\footnote{The set-based fairness used by \citet{yang2017measuring} are defined at increments of 10 rank positions, ranging from 10 to $N$. The measures provide fairness values in the range $[0,1]$, where 0 means most fair and 1 means most unfair. Therefore we refer to this family of metrics as (un)fairness metrics.}
These (un)fairness metrics measure the statistical parity in ranked outputs, i.e., they quantify the relative representation of the protected group \(S^+\) in a ranking \emph{r}.

\begin{table}[tp]
\caption{Notation used in (un)fairness metrics.}
\label{table:fairness_notations_table}
 \begin{tabular}{lp{6cm}} 
 \toprule
 Notation & Description \\ 
 \midrule
 $N$ & Number of relevant arguments in ground truth \\ 

 $r$ & A ranking  \\ 

 \(S^+\) & Protected group \\

  \(S^-\) & Unprotected group \\
\bottomrule
\end{tabular}
\end{table}

\noindent \textbf{Normalized Discounted Difference (\rNDTopk).} \rNDTopk computes the discounted difference between the proportion of arguments belonging to the protected group in the top-$i$ subsets (represented by $\frac{|S^{+}_{1 \ldots i}|}{i}$) and the overall population ($\frac{|S^+|}{N}$):

\begin{equation}
\rNDTopk(\emph{$r$}) = \sum_{i=1}^{\emph{$k$}} 
\frac{1}{log_2(i+1)} \bigg|\frac{|S^{+}_{1 \ldots i}|}{i} - \frac{|S^+|}{N}\bigg|
\label{formula_rND}
\end{equation}

\noindent \textbf{Normalized Discounted KL-divergence (\rKLTopk).} \rKLTopk calculates the Kullback-Leibler (KL) divergence to measure the distance between two probability distributions $P$ and $Q$:
 
 \begin{equation}
 P = \left( \frac{|S^{+}_{1 \ldots i}|}{i},\frac{|S^{-}_{1 \ldots i}|}{i}\right)
 , Q = \left( \frac{|S^{+}|}{N},\frac{|S^{-}|}{N}\right)
 \end{equation}

\noindent where $P$ represents the proportion of the protected group until rank $i$ in top-$i$ and $Q$ represents the proportion of the protected group in the overall population. Intuitively, \rKL measures the difference between the proportion of the protected group in the top-$i$ ranks, and in the overall population:

\begin{equation}
\rKLTopk(\emph{$r$}) = \sum_{i=1}^{\emph{$k$}} \frac{D_{KL}(P||Q)}{log_2(i+1)}
\label{formula_rKL}
\end{equation}

\noindent where $D_{KL}(P||Q)$ is the KL-divergence between $P$ and $Q$.

\noindent \textbf{Normalized Discounted Ratio (\rRDTopk).} \rRDTopk computes the difference between the ratio of arguments belonging to the protected group to those that are not protected until rank $i$, and the ratio of the same in the overall population:

\begin{equation}
\rRDTopk(\emph{$r$}) = \sum_{i=1}^{\emph{$k$}}  
\frac{1}{log_2(i+1)} \bigg|\frac{|S^{+}_{1 \ldots i}|}{|S^{-}_{1 \ldots i}|} - \frac{|S^{+}|}{|S^{-}|}\bigg|
\label{formula_rRD}
\end{equation}
As mentioned by \citet{yang2017measuring}, when either the number or the denominator of a fraction is 0, the entire fraction is set to 0.
After computation we apply min-max normalization to the three (un)fairness metrics.

\subsection{Diversity Metric: $\boldsymbol{\alpha}$-nDCG}

Exposing the different stances of an argument in a ranking can be seen as a search result diversification problem, i.e., rankings with diversified stances. Different diversity metrics have been proposed and analyzed in the literature~\cite{amigo2018axiomatic, sakai2019which}. In our study, we consider \alphandcg proposed by \citet{clarke2008novelty} to analyze the relationship between (un)fairness and diversity.

The parameter $\alpha \in [0,1]$ controls the reward of diversity and novelty in the ranking. When $\alpha=0$, \alphandcg is equivalent to $\ndcg$, and the higher the $\alpha$ value, the more reward is obtained by diversifying the ranking---at the expense of reward obtained by relevance.




%% file: 04-setup.tex
\section{Experimental Setup}

\subsection{Real Scenario}

\noindent \textbf{Test Collection.} In our analysis, we use the test collection used at the \touche Lab at CLEF 2020, comprising 387,740  documents (arguments) from the args.me corpus \cite{ajjour2019data}, and a ground truth of 2,964 
relevance judgments across 49 topics with relevant arguments graded in the range of 1 (least relevant) to 5 (most relevant).

\noindent \textbf{Systems.}  We use the 21 official runs submitted to the \touche argument retrieval task. The systems cover a variety of retrieval, query expansion, and re-ranking techniques~\cite{bondarenko2020overview}. Note that neither fairness nor diversity were considered in the original setup of the challenge, and only topical relevance and argument quality needed to be optimized.

\noindent \textbf{Protected Group.}
In this paper, we make the assumption that, for each topic, the stance with the lowest number of arguments judged as relevant is considered as the protected group.
Protected group in the fairness literature is identified as a group that is disadvantaged/under-represented or a minority. Inspired from this definition, in this paper, we make the assumption that, for each topic, the stance with the lowest number of relevant arguments (i.e., minority) is considered as the protected group.

\subsection{Controlled Scenario}

\begin{figure}[tp]
    \centering
    \includegraphics[width=1\linewidth,trim=0cm 12.5cm 0cm 13.5cm,clip=true]{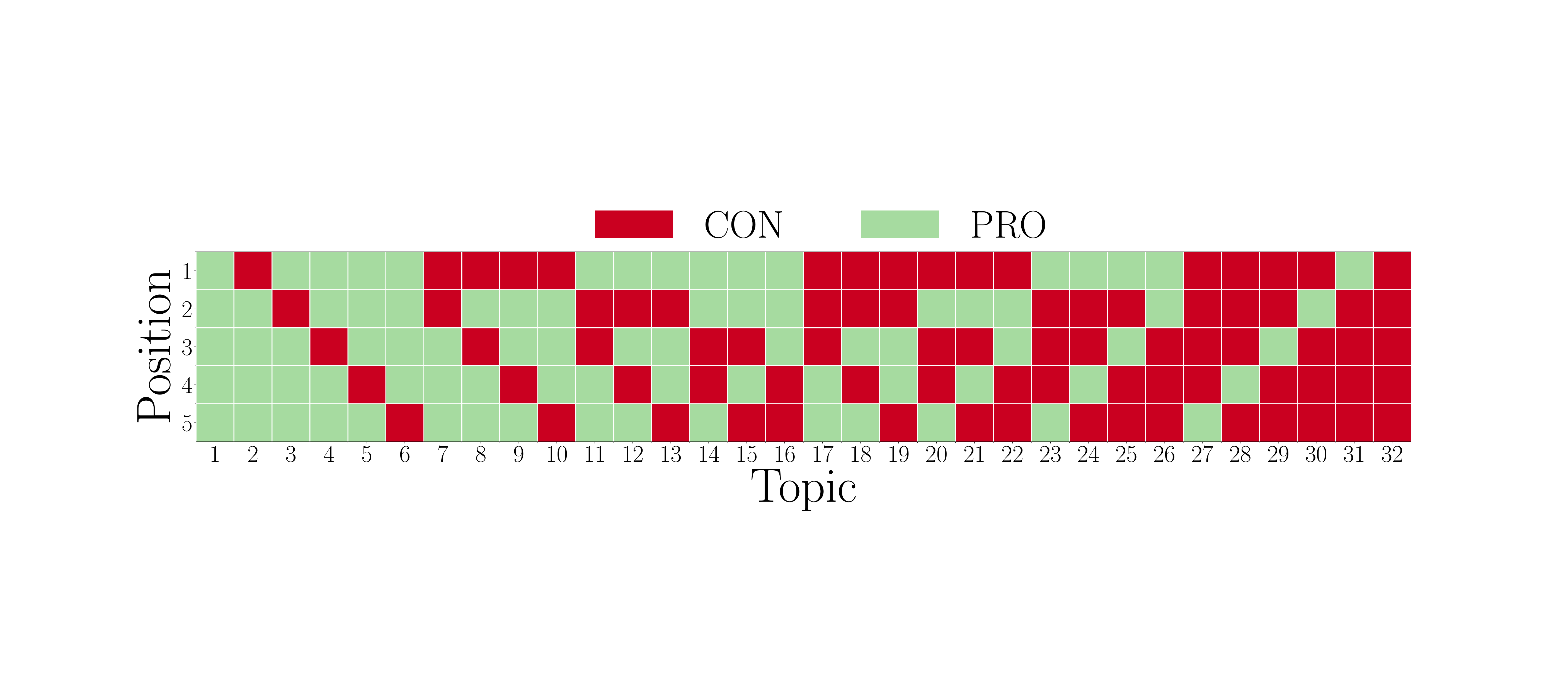}
    \caption{Permutations of argument stances across the top-5 positions in the generated synthetic data.}
    \label{fig:synthetic_data_matrix_generation}
\end{figure}

To systematically study the behaviour of (un)fairness and diversity measures, we additionally created a synthetic test collection which consists of the 32 possible permutations of argument stances and rankings of length 5. Figure ~\ref{fig:synthetic_data_matrix_generation} shows the rankings generated for the controlled scenario. 
For the sake of simplicity, we only consider binary relevance in the synthetic data. Each permutation represents a topic and a ranking of 5 relevant arguments. The number of relevant judgments for each stance across topics vary according to the three definition of protected group described below.

\noindent \textbf{Protected Groups.} We consider three different definitions of protected group.
\begin{itemize}
    \item \emph{Minority Group.} For all topics, the proportion of \pro/\con arguments judged as relevant has a 4/1 ratio. The minority group (\con) is considered as protected \(S^+\).\footnote{Note that, as the distribution of \pro and \con across topics is symmetric,  a ratio of 1/4 and choosing \pro as protected group would lead to the same results.}
    \item \emph{Proportion-Agnostic.} All topics have a 1/1 ratio of \pro/\con arguments judged as relevant in the ground truth. The protected group is then defined independently of the the proportion (e.g., a possible scenario would be that this decision is informed by a policy). We set \pro as the protected group, although choosing one stance or the other would lead to the same outcome in this case.
    \item \emph{Majority Group.} For all topics, the proportion of \pro/\con arguments judged as relevant has a 4/1 ratio. The majority group (\pro) is considered as protected \(S^+\). 
\end{itemize}

\subsection{Evaluation Metrics}
In our analysis we consider the official effectiveness evaluation of the \touche evaluation campaign, which only considers relevance, \ndcgTopk. Scores were computed using \texttt{trec\_eval}.
For (un)fairness metrics, we computed \rND, \rKL, and \rRD using our own implementation.\footnote{The source code is available at \url{https://github.com/sachinpc1993/fair-arguments}.}
To compute the diversity metric \alphandcgTopk, we used \texttt{ndeval} from the TREC 2014 Web Track.\footnote{\url{https://github.com/trec-web/trec-web-2014/blob/master/src/eval/ndeval.c}} Unless otherwise specified, we use the default parameter value $\alpha=0.5$.

\subsection{Meta-Evaluation}
In order to analyze changes on ranking of systems, we compare evaluation measures using Kendall's $\tau$ correlation \cite{kendall1938new}.\footnote{We are aware of alternative ways of comparing rankings \cite{yilmaz2008new, kutlu2018when, ferro2017what, carterette2009rank}. As we are not defining an absolute threshold in correlations scores -- and for the sake of interpretability of the results -- we chose Kendall's $\tau$ to compare evaluation measures.} 
Kendall's $\tau$ provides an intuitive way to compare the rankings of systems, i.e., $\frac{1-\tau}{2}$\% of system pairs are those with a different relative ordering in the two rankings. $\tau$ values range between $[-1,1]$ and does not consider the magnitude of the item values but only their relative ordering. For the real scenario, scores are averaged across topics and system rankings are compared. For the controlled scenario, correlations are computed using the scores obtained for synthetic rankings.

%% file: 05-results.tex
\section{Results and Analysis}
\label{sec:results}

\subsection{Real Scenario: \touche Lab at CLEF 2020}

\begin{figure}[tb]
    \centering
    \includegraphics[width=1\linewidth]{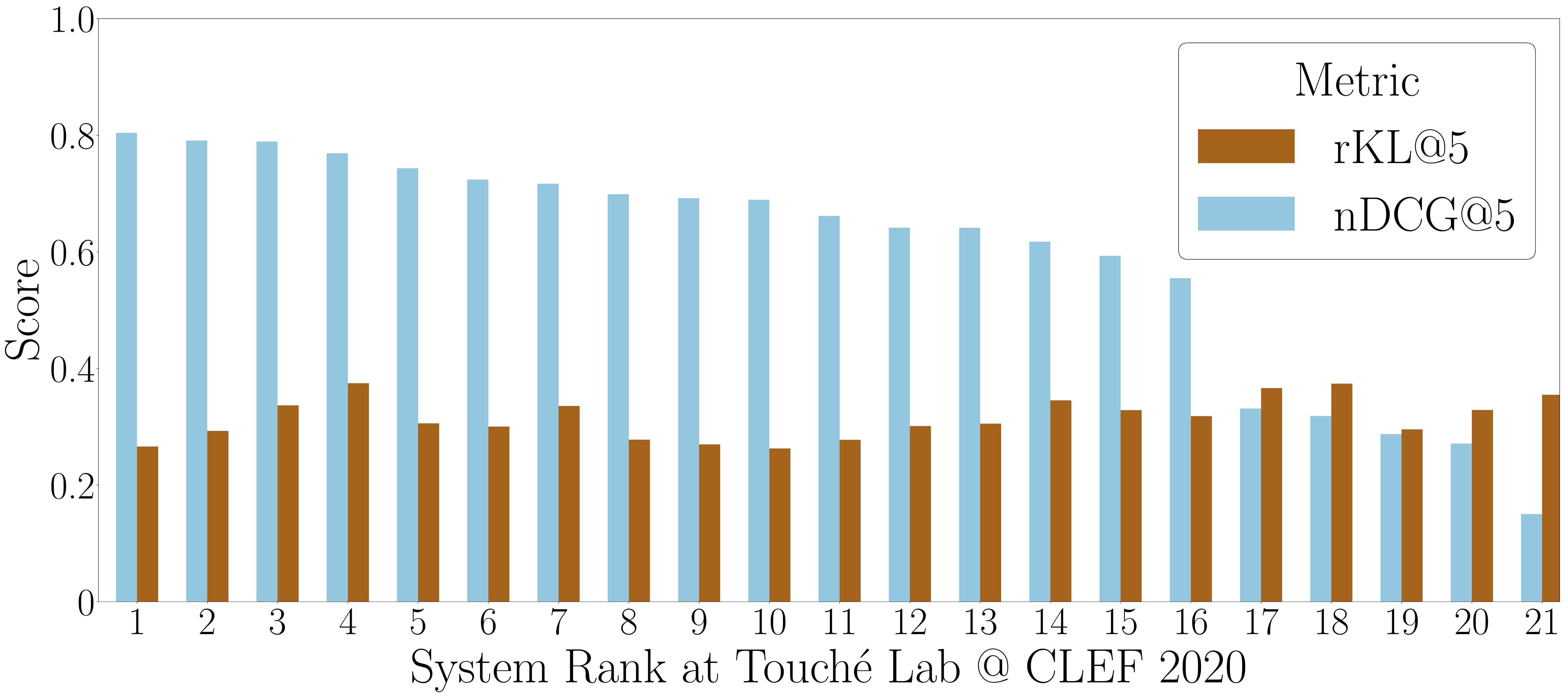}
    \caption{\ndcgTopkCap (higher is better) and \rKLCap (un)fairness (lower is better) for official runs of the \touche argument retrieval task at CLEF 2020. Systems (x-axis) are shown in decreasing order of \ndcgTopkCap.}
    \label{fig:ranking_order_rKL_ndcg}
\end{figure}

The effectiveness of runs submitted to \touche are shown in Figure~\ref{fig:ranking_order_rKL_ndcg},
sorted by the official evaluation measure of the challenge: \ndcgTopk. 
Comparing \ndcgTopk scores with \rKL we can see that \rKL scores do not increase monotonically. The same trend was observed for \rND and \rRD. This suggests that the system rankings would be altered if both relevance and (un)fairness metrics are considered. However, it is worth noting that some of the ranked systems at \touche would remain unchanged, e.g., the top-3 systems are the same if the harmonic mean of \alphandcgTopk and $1-$\rKL is used to rank the systems.

\begin{table}[tb]
\caption{Kendall's $\mathbf{\boldsymbol\tau}$ correlation between (un)fairness metrics, \ndcgTopkCap, and \alphandcgTopkCap for the \touche runs (real scenario). * indicates statistical significance $\mathbf{(p<0.05)}$. \label{tab:real_scenario_tau_correlations}
}
\begin{adjustbox}{max width=1\linewidth}
\sisetup{round-mode=places,round-precision=4}
 \begin{tabular}{c
                 S[table-format=1.4]
                 S[table-format=1.4]
                 S[table-format=1.4]
                 S[table-format=1.4]
                 } 
 \toprule
 (un)fairness & \ndcgTopk  & \alphandcgTopk & \rND                      & \rKL                                                                                                                             \\
 \midrule
 \rND         & -0.0762    &  -0.0667       &  
                                                                        &                                    \\
 \rKL         & -0.2667    &  -0.3333*      &  0.3523809523809524*     &    
                                                                                                          \\
 \rRD         & -0.2571    &  -0.3238*      &  -0.1142857142857143     &    0.2857142857142857             
                                                                                                            \\
 \bottomrule
\end{tabular}
\end{adjustbox}
\end{table}

The Kendall's $\tau$ correlations between the relevance, (un)fairness, and diversity metrics considered in our study are shown in Table~\ref{tab:real_scenario_tau_correlations}.
As expected, the (un)fairness metrics are negatively correlated with the relevance-based metric \ndcgTopk (optimal scores are 0 and 1, respectively). The correlations are low and not statistically significant.
\alphandcgTopk is more similar to \rKL and \rRD, although the correlations are still weak. This suggests that while novelty and diversity are partially captured by these (un)fairness metrics, they still measure different dimensions. We also observed low correlations between \rND and the other metrics.

Meta-evaluation of the \touche challenge shows that: (i) diversity is related but not equivalent to (un)fairness; (ii)  (un)fairness metrics lead to different system rankings; and (iii) (un)fairness metrics therefore provide complementary information to relevance-based metrics, and considering both would lead to changes in the leaderboard.
From the real scenario, where different topics have different proportions of \pro/\con stances, it is not clear whether a particular (un)fairness metric is more appropriate than another. To understand how the positioning of the argument stance can affect the behavior of diversity and (un)fairness metrics, we next analyze these metrics in a controlled scenario with synthetic data.  

\subsection{Controlled Scenario: Synthetic Data}

Kendall's $\tau$ correlations between (un)fairness metrics and \alphandcgTopk are shown in Table~\ref{tab:conrelations_controlled_alpha_05_diversity_fairness} for three different definitions of protected groups. Here, we compare metrics over all possible permutations of depth 5 argument rankings -- with all arguments in the rankings being equally relevant. 
The following observations can be made. 
\rND has a higher negative correlation with \alphandcgTopk for the \emph{Minority} setting, but not for the other settings.
The symmetric treatment of the groups \(S^+\) and \(S^-\) in the \rKL measure is reflected in the correlations scores for the \emph{Minority} and \emph{Majority} settings.

It is also worth noting that \alphandcgTopk scores for the three settings
are the same: this suggests that, if the definition of protected group is based on proportion of the population (e.g., statistical parity), \alphandcgTopk cannot be used.
When \alphandcgTopk is calculated using  $\alpha=0.9$ (Table~\ref{alpha_09_diversity_fairness}), higher negative correlations are obtained for \rKL, but not for \rND and \rRD. 
\rRD has the same correlation with \alphandcgTopk for different values of $\alpha$ (in both \emph{Minority} and \emph{Majority} settings), while the correlation is substantially lower for the \emph{Proportion-Agnostic} setting.

\begin{table}[tb]
\sisetup{round-mode=places,round-precision=4}
\caption{Kendall's $\boldsymbol\tau$ correlation between (un)fairness metrics and \alphandcgTopkCap diversity metric for different definitions of protected groups in the synthetic data (controlled scenario). * indicates statistical significance $\mathbf{(p<0.05)}$.
}
\label{tab:conrelations_controlled_alpha_05_diversity_fairness}
\begin{adjustbox}{max width=1\linewidth}

 \begin{tabular}{c
 S[table-format=-1.4]
 S[table-format=-1.4]
 S[table-format=-1.4]
 } 
 \toprule
  & \multicolumn{3}{c}{Protected Group \(S^+\)}\\
 (un)fairness & {Minority} & {Proportion-Agnostic} & {Majority}\\ 
 \midrule
 \rND & -0.4812* & -0.1434 & 0.0\\
  \rKL & -0.1747 & -0.8378* & -0.1747\\ 
 \rRD & 0.1032 & -0.6924* & -0.2572\\
 \bottomrule
\end{tabular}
\end{adjustbox}
\vspace{-1.5ex}
\end{table}

\begin{table}[tb]
\caption{Kendall's $\boldsymbol\tau$ correlation between (un)fairness metrics and \alphandcgTopkCap diversity metric with \(\boldsymbol{\alpha = 0.9} \). * indicates statistical significance $\mathbf{(p<0.05)}$.}
\label{alpha_09_diversity_fairness}
\begin{adjustbox}{max width=1\linewidth}

 \sisetup{round-mode=places,round-precision=4}
 \begin{tabular}{c
  S[table-format=-1.4]
 S[table-format=-1.4]
 S[table-format=-1.4]}
  \toprule
  & \multicolumn{3}{c}{Protected Group \(S^+\)} \\
 (un)fairness & {Minority} & {Proportion-Agnostic} & {Majority} \\ 
 
\midrule
 \rND & -0.4554* & -0.1520 & 0.0\\

 \rKL & -0.2003 & -1.0* & -0.2003\\ 

 \rRD & 0.1032 & -0.5799* & -0.2572\\
 \bottomrule
\end{tabular}
\end{adjustbox}
\vspace{-1.7ex}
\end{table}

A further examination of the individual topic scores showed that, for extreme cases where all the top-5 arguments belong to either \(S^+\) or \(S^-\) (i.e., topics 1 and 32 in Figure~\ref{fig:synthetic_data_matrix_generation}), 
\rKL does not respond as one would expect. For instance, in the \emph{Minority} setting, \rKL shows low (un)fairness (0.1463) for the topics that has all arguments belonging to \(S^-\), and high (un)fairness (1.0000) for the topic that has all \(S^+\). \rND shows similar but less skewed behavior, and the difference in \rND scores for symmetric topics is substantially lower than is observed for \rKL. On the other hand, \rRD gives high (un)fairness scores for the extreme cases.

Our analysis in a controlled scenario shows that, although diversification in terms of argument stances may lead to a fair exposure of perspectives, diversity and (un)fairness metrics are not necessarily measuring the same dimensions.

%% file: 06-conclusions.tex
\section{Conclusion and Future Work}

\label{sec:conclusion}

To our knowledge, the notion of fair exposure of different stances in argument retrieval systems have not been explored before. We analyze the relation between relevance, group fairness, and diversity measures in the context of argument retrieval. 
Our analysis of the systems submitted to the \touche argument retrieval evaluation campaign shows that the outputs from argument retrieval systems can be biased towards a particular stance, and system rankings change when (un)fairness metrics are considered. Our correlation analysis using synthetic data corroborates that fairness-aware rankings do not guarantee diverse results and vice-versa. 

In this work, we considered only a binary protected attribute (\pro and \con). Given that Normalized Discounted KL-divergence is a fairness measure capable of handling multi-level protected attributes, it is worth exploring how the complexity of such protected attributes affects the relationship between (un)fairness and diversity in future work. Stochastic fairness metrics~\cite{diaz2020evaluating} or other diversity measures~\cite{gao2020toward,amigo2018axiomatic,sakai2019which} are yet to be investigated for argument retrieval. However, this requires re-purposing the existing test collections.